\documentclass{iopart}

\usepackage{graphicx}
\usepackage{bm}
\usepackage{sidecap}

\begin{document}

\title{The 4-spinon dynamical structure factor of the Heisenberg chain}

\author{
Jean-S\'ebastien Caux and
Rob Hagemans
}
\address{Institute for Theoretical Physics, University of
  Amsterdam, 
  1018 XE Amsterdam, The Netherlands.}

\date{\today}

\begin{abstract}
We compute the exact 4-spinon contribution to the zero-temperature dynamical structure factor of the
spin-$1/2$ Heisenberg isotropic antiferromagnet in zero magnetic field, directly 
in the thermodynamic limit.  We make use of the expressions for matrix elements
of local spin operators obtained by Jimbo and Miwa using the quantum affine symmetry
of the model, and
of their adaptation to the isotropic case by Abada, Bougourzi and Si-Lakhal (correcting
some overall factors).  The 4-spinon contribution to the first frequency moment sum rule at
fixed momentum is calculated.  This shows, as expected, that most of the remaining 
correlation weight above the known 2-spinon part is carried by 4-spinon states.
Our results therefore provide an extremely accurate description of the exact structure factor.
\end{abstract}

\maketitle

\section{Introduction}
It has now been 75 years since Hans Bethe published his seminal paper \cite{BetheZP71}
constructing the eigenfunctions of the spin-$1/2$ Heisenberg chain \cite{HeisenbergZP49},
\begin{equation}
H = \sum_{j=1}^N {\bf S}_j \cdot {\bf S}_{j+1},
\label{XXX}
\end{equation}
giving birth to the Bethe Ansatz and paving the way for the modern theory of integrable models of quantum mechanics
and field theory \cite{KorepinBOOK,TakahashiBOOK,EsslerREVIEW}.  Interest in the Heisenberg model has only increased since those
early days, partly because of the extremely rich mathematical structures now known to
be associated to it, but also because of its ability to accurately describe
a number of real compounds.

The Bethe Ansatz is first and foremost a method giving access to an integrable
system's energy levels, allowing the calculation of many equilibrium quantities.  
For the specific case of the Heisenberg model, the ground state energy of the infinite chain was
computed analytically \cite{HulthenAMAFA11} not long after Bethe's paper, 
but it was not until the 1960's that significant new results were obtained:  
its excitation spectrum was computed by des Cloizeaux and Pearson \cite{desCloizeauxPR128},
and its more general thermodynamic properties were obtained shortly afterwards 
\cite{GriffithsPR133,YangPR150,GaudinPRL26,TakahashiPTP46_1}.

Equilibrium quantities are however not sufficient to completely characterize the
physics of models such as (\ref{XXX}).  Motivated mainly by experimental work, another
object of fundamental importance has been extensively studied:  the dynamical structure factor
(DSF)
\begin{equation}
S^{ab} (k, \omega) = \sum_{l=1}^N e^{i k l} \int_{-\infty}^{\infty} dt e^{i\omega t} \langle S_{j+l}^a (t) S_j^b(0) \rangle
, \hspace{1cm}\mbox{a,b = x, y, z}.
\label{S_k_omega}
\end{equation}
This quantity is directly accessible experimentally through inelastic neutron scattering
\cite{KenzelmannPRB65,StonePRL91,ZaliznyakPRL93,LakeNature4}, 
and its theoretical calculation opens the door to the interpretation of a wealth of
experimental data.

Despite much effort, the dynamics of integrable models remains in general inaccessible, and 
exact calculations based on the Bethe Ansatz can only rarely be carried out.  One reason is
that excitations are usually rather complicated:  for the Heisenberg model, the Hilbert space can
be spanned with spinons \cite{FaddeevPLA85,FowlerPRB18,AndersonSCIENCE235}, which are nontrivially
interacting spin-$1/2$ particles whose dispersion relation is given by
\begin{equation}
e (p) = \frac{\pi}{2} |\sin p|, \hspace{1cm} p \in [-\pi, 0].
\label{spinon_disp_rel}
\end{equation}
The computation of dynamical quantities such
as the structure factor however requires knowledge of matrix elements of spin operators 
between (multi-)spinon states, which goes much beyond what is accessible with the basic Bethe Ansatz.

In view of the difficulty of this task, a number of approximate schemes have been put forward
to offer a qualitative picture of the DSF.  One extremely useful construction is known as the
M\"uller ansatz \cite{MuellerPRB24}, which is based on exact results for the $XY$ model, numerical
computations on small chains and known sum rules.  Its success lies in its extreme simplicity, coupled
with rather accurate reproduction of a number of features (like the square-root singularity at the
lower threshold).  It is commonly used in the interpretation of experimental data.  It drawback
is that it is inexact;  in particular, its functional form at the top of the 2-spinon continuum
is not correct.

Another important and successful approach relies on mapping the infinite chain onto a relativistic
quantum field theory \cite{LutherPRB9,KadanoffAP121}.  Finite size scaling connects the critial
exponents of the system with its behaviour in a finite volume \cite{AffleckPRL56,BloetePRL56},
while conformal field theory \cite{BelavinNPB241,CFTBOOK} and bosonization allow the calculation of asymptotics of correlation
functions \cite{AffleckLESHOUCHES,BosonizationBOOK,GiamarchiBOOK} even at finite temperatures, with known
normalizations for the first few leading terms in the operator expansion \cite{AffleckJPA31,LukyanovPRB59,LukyanovNPB654}.

As far as methods based on integrability are concerned, it is now possible to achieve extremely
accurate computations of the DSF over the whole Brillouin zone \cite{CauxPRL95,CauxJSTATP09003}
for integrable Heisenberg spin chains of any anisotropy in any magnetic field, using the ABACUS method \cite{ABACUS}.
These computations rely on determinant representations for matrix elements of local spin operators
obtained by solving the quantum inverse problem \cite{KitanineNPB554,KitanineNPB567}, and are
therefore limited to finite (albeit large) chains.  We here wish however to concentrate on an
altogether different and independent take on the problem of calculating the DSF of the Heisenberg model, which
is not applicable in general but only in one particular (albeit extremely important) case.

The crucial development (as far as the subject of the present paper is concerned) came with the recognition
that the infinite chain in zero magnetic field displayed a quantum affine symmetry, allowing the diagonalization of
the Hamiltonian directly in the thermodynamic limit \cite{DaviesCMP151}.  
Multi-spinon states then provide a basis for the Hilbert space and a 
resolution of the identity operator, allowing to write the DSF as a sum of matrix elements of local
spin operators.  These matrix elements can be computed within this framework through bosonization of
the quantum affine algebra \cite{FrenkelPNAS85,AbadaMPLA8,BougourziNPB404}, a task performed by
Jimbo and Miwa \cite{JimboBOOK}.  The general representation of such correlation functions as the DSF is then
obtained in terms of rather complicated contour integrals, the number of integrals increasing with 
the number of spinons in the excited state.  These are however notoriously hard to
evaluate quantitatively.  

One exception is the contribution to the transverse zero-temperature DSF of the Heisenberg model 
coming from 2-spinon intermediate states, the
simplest excitations that can be constructed above the ground state.  In this case, the dynamical
constraints of conservation of energy and momentum give two $\delta$ functions taking care of 
the two contour integrals involved, allowing for a direct analytical calculation \cite{BougourziPRB54} (a similar
calculation is also possible in the whole gapped antiferromagnetic phase \cite{BougourziPRB57}).  
These 2-spinon intermediate states were shown to
contribute $72.89\%$ of the total structure factor intensity \cite{KarbachPRB55}.  The missing part
is then necessarily carried by excited states with a higher number of spinons, starting with
4-spinon states.

The integral representations for matrix elements involving 4-spinon states are much more
difficult to tackle, since the dynamical constraints are not sufficient to take care of all
the contour integrals involved (this is also true for the longitudinal structure factor of the $XXZ$
model, which was studied in the Ising limit in \cite{WestonCRM}).  The first attempt to tackle
more than two spinons was made in \cite{BougourziMPLB10}, offering a formal representation for
the exact $n$-spinon contribution for the zero-temperature DSF of the Heisenberg model in zero
field.  A more thorough treatment of the four-spinon case was published shortly afterwards \cite{AbadaNPB497}, 
yielding expressions for the DSF in the whole gapped antiferromagnetic regime, and their
specialization to the Ising and Heisenberg limits.  These expressions remained however
quite complicated, giving the four-spinon part at fixed momentum and energy in terms of 
a double integral of an infinite series.  Further analytical work \cite{Abada9802271} yielded
little progress, and in fact (as we will show below) incorrectly identified the boundaries of the 
four-spinon continuum.  To this day, nobody has been able to extract
curves from these expressions, and previous attempts \cite{AbadaJPA37,SiLakhalPB369}
have not yielded acceptable results due to the inappropriateness of the chosen method
and the incorrect continuum used.

The present paper offers the first reliable computation of the exact 4-spinon contribution to the
zero-temperature dynamical structure factor of the Heisenberg isotropic antiferromagnet in zero
magnetic field.  After summarizing results for the
matrix elements which we need for our purposes, the known 2-spinon results are
repeated as a warmup, after which we present their extension to the case of four spinons 
(correcting some factors in the formulas present in the literature).  We finish by a
discussion of our results, in particular concerning contributions to sum rules.

\section{Exact representation of the dynamical structure factor}

The inevitable first step in the calculation of the DSF (\ref{S_k_omega}) is to insert
a resolution of the identity in a judiciously chosen basis between the two spin operators.
For the $XXZ$ model in the thermodynamic limit, a basis for the Hilbert space is provided by
(multi-) spinon states $| \xi_1, ..., \xi_n \rangle_{\epsilon_1, ... ,\epsilon_n;i}$,
which diagonalize the Hamiltonian according to
\begin{eqnarray}
H | \xi_1, ..., \xi_n \rangle_{\epsilon_1, ... ,\epsilon_n;i} = \sum_{j=1}^n e (\xi_j) 
| \xi_1, ..., \xi_n \rangle_{\epsilon_1, ... ,\epsilon_n;i}
\end{eqnarray}
where $i$ labels the two equivalent vacuum states $|0\rangle_i$, $i = 0, 1$ corresponding to the
two different possible boundary conditions, and $\epsilon_i = \pm 1$ labels the spin projection
of the spinons.  The coefficients $\xi_j$ are spectral parameters determining the energy and momenta of 
the spinons.  The completeness relation reads
\begin{equation}
\fl
{\bf 1} = \sum_{i=0,1} \sum_{n \geq 0} \sum_{\epsilon_1,...,\epsilon_n = \pm 1}
\frac{1}{n!} \oint \prod_{j=1}^n \frac{d\xi_j}{2\pi i \xi_j} | \xi_n, ..., \xi_1 \rangle_{\epsilon_n, ... ,\epsilon_1;i}
{~}_{i; \epsilon_1, ... ,\epsilon_n} \langle \xi_1, ..., \xi_n |
\label{completeness}
\end{equation}
and can be substituted in (\ref{S_k_omega}) to yield the decomposition of the DSF into
a sum over (even numbers of) spinon contributions
(here and in the following,
we make use of spin isotropy and compute $S(k, \omega) \equiv S^{zz} (k, \omega) = S^{xx} = S^{yy}$)
\begin{eqnarray}
S (k, \omega) = \sum_{n ~\small{even}} S_n (k, \omega).
\end{eqnarray}
Each term in this decomposition is explicitly written as
\begin{eqnarray}
\fl S_n (k, \omega) = \frac{2\pi}{n!} \sum_{m \in \mathbf{Z}} \sum_{\epsilon_1,...,\epsilon_n = \pm 1}
\oint \prod_{j=1}^n \frac{d\xi_j}{2\pi i \xi_j} e^{i m (k + \sum_{j=1}^n p_j)} 
\delta (\omega - \sum_{j=1}^n e_j) \times \nonumber \\
\hspace{1cm}\times 
{}_i\langle| 0| S^z_0 (0) | \xi_n, ..., \xi_1 \rangle_{\epsilon_n, ... ,\epsilon_1;i}
{~}_{i; \epsilon_1, ... ,\epsilon_n} \langle \xi_1, ..., \xi_n | S^z_0(0) | 0 \rangle_i.
\end{eqnarray}
The matrix elements of local spin operators in the above expression are
represented exactly within the framework of Jimbo and Miwa \cite{JimboBOOK}.
Their adaptation to the isotropic Heisenberg antiferromagnet was given in
\cite{BougourziPRB54} for 2-spinon intermediate states, and in \cite{BougourziMPLB10}
for $n > 2$, although the expressions obtained there are not thoroughly simplified.
For 4-spinon intermediate states, the matrix elements were studied more extensively
in \cite{AbadaNPB497}, whose results will form the basis of the new results we
obtain.  Let us however start by briefly reminding the reader of known results on the much simpler case
of 2-spinon intermediate states.  

\section{2-spinon contribution to the structure factor}
It is well-known that 2-spinon intermediate states live within a continuum in $k, \omega$ defined by 
satisfying the kinematic constraints of momentum and energy conservation, 
\begin{equation}
k = -p_1 - p_2, \hspace{1cm} \omega = e (p_1) + e (p_2).
\end{equation}
In other words, for a fixed external momentum, there exists an interval in frequency
given by the conditions
\begin{equation}
\fl 
\omega \geq \omega_{2,l} (k) = \frac{\pi}{2} |\sin k|, \hspace{1cm}
\omega \leq \omega_{2,u} (k) = \pi \sin \frac{k}{2}, \hspace{1cm} k \in [0, 2\pi].
\label{2p_continuum}
\end{equation}
The lower boundary is thus given by the des Cloizeaux-Pearson dispersion relation.  
The 2-spinon part of the DSF will be nonvanishing within this continuum, and will by construction vanish
identically outside of it.  This contribution was
obtained in \cite{BougourziPRB54} and is explicitly written as
\begin{equation}
S_2 (k, \omega) = \frac{1}{2} \frac{e^{-I (\rho(k,\omega))}}{\sqrt{\omega_{2,u}^2(k) - \omega^2}} 
\Theta (\omega_{2,u} (k) - \omega) \Theta (\omega - \omega_{2,l}(k)).
\label{S_2}
\end{equation}
The parameter $\rho$ is defined as
\begin{equation}
\cosh (\pi \rho(k, \omega)) = \sqrt{\frac{\omega_{2,u}^2(k) - \omega_{2,l}^2(k)}{\omega^2 - \omega_{2,l}^2(k)}}
\end{equation}
and the nontrivial part of the DSF is encoded in the fundamental integral function
\begin{equation}
I (\rho) = \int_0^{\infty} dt \frac{e^t}{t} \frac{\cosh (2t) \cos (4\rho t) - 1}{\cosh t \sinh (2t)}.
\label{I_rho}
\end{equation}

A careful study of this representation of the 2-spinon DSF was carried out in \cite{KarbachPRB55}, and it is 
worthwhile to remind the reader of some important facts obtained there.  
First of all, at the lower boundary (for $q \neq \pi$), the 2-spinon DSF diverges as a square root 
accompanied by a logarithmic correction, $S_2 \sim \frac{1}{\sqrt{\omega - \omega_{2,l}(k)}} \sqrt{\ln \frac{1}{\omega - \omega_{2,l}(k)}}$ 
(for $q = \pi$, the divergence is $\sim \frac{1}{\omega} 
\sqrt{\ln \frac{1}{\omega}}$).  Near the upper boundary, on the other hand, the 2-spinon
DSF vanishes in a square-root cusp, $S_2 \sim \sqrt{\omega_{2,u}(k) - \omega}$.

Second, sum rules were also studied, most importantly the contribution to the total integrated intensity
\begin{equation}
\int_0^{2\pi} \frac{dk}{2\pi} \int_0^{\infty} \frac{d\omega}{2\pi} S(k, \omega) = \frac{1}{4}
\end{equation}
and to the exactly known first frequency moment at fixed momentum,
\begin{equation}
K_1 (k) = \int_0^{\infty} \frac{d\omega}{2\pi} \omega S(k, \omega) = (1 - \cos k) \frac{2 e_0}{3}
\label{K_1}
\end{equation}
where $e_0 = 1/4 - \ln 2$ is the ground-state energy density \cite{HulthenAMAFA11}.
2-spinon intermediate states were shown to carry $72.89\%$ of the total intensity,
and $71.30\%$ of the first moment sum rule (independently of $k$).  

More than a quarter of the exact DSF is therefore missing if we restrict ourselves to 
only two spinons.  To achieve better saturation of the sum rules, we need to go to more
complicated intermediate states involving more particles, and we can safely expect that
out of those, 4-spinon states will be dominant.

\section{4-spinon contribution to the structure factor}
Starting from the results of \cite{AbadaNPB497}, we write the exact representation for the
4-spinon part of the DSF as
\begin{equation}
\fl
S_4 (k, \omega) = C_4 \int_{-\pi}^0 dp_1 ... dp_4 ~~\delta_{(2\pi)} (k + \sum_i p_i) 
~~\delta (\omega + \sum_i e (p_i)) ~~J (\{ p \})
\label{S_4_1}
\end{equation}
where the prefactor is
\begin{eqnarray}
C_4 = \frac{1}{3 \times 2^9} \frac{1}{\Gamma(1/4)^8 |A(i \pi/2)|^8},
\end{eqnarray}
with
\begin{equation}
A(z) = \exp \left( - \int_0^{\infty} dt \frac{e^t}{t} \frac{\sinh^2 (t[1 + i \frac{z}{\pi}])}{\sinh (2t) \cosh t}\right).
\end{equation}
Here again, we restrict to $k \in [0, \pi]$.  Parametrizing the momenta as
\begin{equation}
\cot p_i = \sinh (2\pi \rho_i)
\end{equation}
the correlation weight is explicitly given by
\begin{equation}
J (\{ p \}) \equiv J(\{ \rho \}) = e^{- \sum_{1 \leq i < j \leq 4} I(\rho_{ij})} \sum_{l=1}^4 |g_l (\{ \rho \})|^2
\end{equation}
where $\rho_{ij} = \rho_i - \rho_j$.  The function $I (\rho)$ is given by equation (\ref{I_rho}), 
whereas $g_l$ is given by the following expression:
\begin{eqnarray}
\fl
g_l = (-1)^{l + 1} \sum_{j=1}^4 \cosh (2\pi \rho_j) \times \nonumber \\
\times \sum_{m = \Theta(j - l)}^{\infty} \frac{\prod_{i \neq l} (m - \frac{1}{2} \Theta(l - i) + i\rho_{ji})}
{\prod_{i \neq j} \sinh(\pi \rho_{ji})} \prod_{i=1}^4 \frac{\Gamma(m - \frac{1}{2} + i\rho_{ji})}{\Gamma (m + 1 + i \rho_{ji})}
\label{g_l}
\end{eqnarray} 
where the Heaviside function is here defined as $\Theta (n) = 0$ for $n \leq 0$ and $\Theta(n) = 1$ for $n > 0$.

We have corrected two inaccuracies in \cite{AbadaNPB497}:  first, the correct normalization is
presented here (compare (\ref{g_l}) with formula (5.10) there), and most importantly, we have
explicitly written that the momentum $\delta$ function fixes $k$ only modulo $2\pi$.  This has the
crucial consequence that two sectors must be considered when solving the dynamical constraints
of momentum and energy conservation for 4-spinon intermediate states (this was overlooked
in \cite{Abada9802271,AbadaJPA37,SiLakhalPB369}, leading in particular to an incorrect description of the
four-spinon continuum).  
Bearing in mind that the spinon momenta $p_i$ are by definition constrained to the interval $[-\pi, 0]$,
we define sectors 0 and 1 as
\begin{equation}
\fl
0:  k + p_1 + p_2 + p_3 + p_4 = 0, \hspace{1cm} 1:  k + 2\pi + p_1 + p_2 + p_3 + p_4 = 0
\end{equation}
with in both cases the energy constraint explicitly written as
\begin{equation}
\omega + \frac{\pi}{2} (\sin p_1 + \sin p_2 + \sin p_3 + \sin p_4) = 0.
\end{equation}
For higher spinon numbers, more sectors must be similarly added:  there are $n$ such sectors
for states with $2n$ spinons.  

A more physical representation of the four-spinon part (\ref{S_4_1}) of the structure factor is 
obtained by the change of variables $\{ p_i \} \rightarrow \{ k, \omega, K, \Omega \}$
where 
\begin{equation}
K = -p_1 - p_2, \hspace{1cm} \Omega = -\frac{\pi}{2} (\sin p_1 + \sin p_2).
\end{equation}
In sector 0, we then have
\begin{equation}
K = k + p_3 + p_4, \hspace{1cm} \Omega = \omega + \frac{\pi}{2} (\sin p_3 + \sin p_4).
\label{KOmega_S0}
\end{equation}
The complete transformation reads
\begin{eqnarray}
\fl
&p_1 = -\frac{K}{2} + \mbox{acos} \frac{\Omega}{\omega_{2,u}(K)}, \hspace{1cm}
&p_2 = -\frac{K}{2} - \mbox{acos} \frac{\Omega}{\omega_{2,u}(K)}, \nonumber \\
\fl
&p_3 = \frac{K - k}{2} + \mbox{acos} \frac{\omega - \Omega}{\omega_{2,u} (k - K)}, \hspace{1cm}
&p_4 = \frac{K - k}{2} - \mbox{acos} \frac{\omega - \Omega}{\omega_{2,u} (k - K)},
\end{eqnarray}
where we restrict to $p_1 > p_2$ and $p_3 > p_4$ by symmetry.  This sector corresponds
to $K \in [0, k]$.

In sector 1, we have 
\begin{equation}
K = k + 2\pi + p_3 + p_4
\end{equation}
instead of the left of (\ref{KOmega_S0}), yielding 
the same expressions for $p_1$ and $p_2$, and
\begin{equation}
\fl
p_3 = -\pi + \frac{K - k}{2} + \mbox{acos} \frac{\omega - \Omega}{\omega_{2,u} (K - k)}, \hspace{0.5cm}
p_4 = -\pi + \frac{K - k}{2} - \mbox{acos} \frac{\omega - \Omega}{\omega_{2,u} (K - k)},
\end{equation}
with the same restrictions as above.  This sector corresponds to $K \in [k, 2\pi]$.  

The 4-spinon continuum in the $k, \omega$ plane is obtained by letting $K$ and $\Omega$
take on all their allowed values.  The lower boundary coincides with that of the 2-spinon
continuum (the des Cloizeaux-Pearson dispersion relation;  
this is clearly the case from a simple physical argument, namely that we are 
dealing with a massless theory, and therefore adding two more spinons of zero momentum in the intermediate
state does not shift the energy;  in fact, the lower boundary of {\it any} finite higher spinon number
continuum is identical to the 2-spinon one).  The upper boundary, on the other hand, extends above the
upper boundary of the 2-spinon continuum, and is obtained by sharing the momentum (modulo $2\pi$) evenly
among the four spinons.  Explicitly, we therefore have
\begin{equation}
\fl
\omega_{4,l} (k) = \omega_{2,l} (k) = \frac{\pi}{2} |\sin k|, \hspace{1cm}
\omega_{4,u} (k) = \pi \sqrt{2\left(1 + |\cos \frac{k}{2}| \right)}.
\end{equation} 
A geometrical picture of the $2n$ spinon continuum is easily obtained by generalization: 
the lower boundary will always be given by the des Cloizeaux-Pearson dispersion relation, 
whereas the upper boundary will be given by the upper boundary of a 2-spinon continuum 
rescaled in size by a factor of $n$, modulo all its $2\pi$ translations (refer to Figure \ref{branch_fig}
for the simplest case of 4 spinons).

\begin{figure}
\begin{tabular}{cc}
\includegraphics[width=6cm]{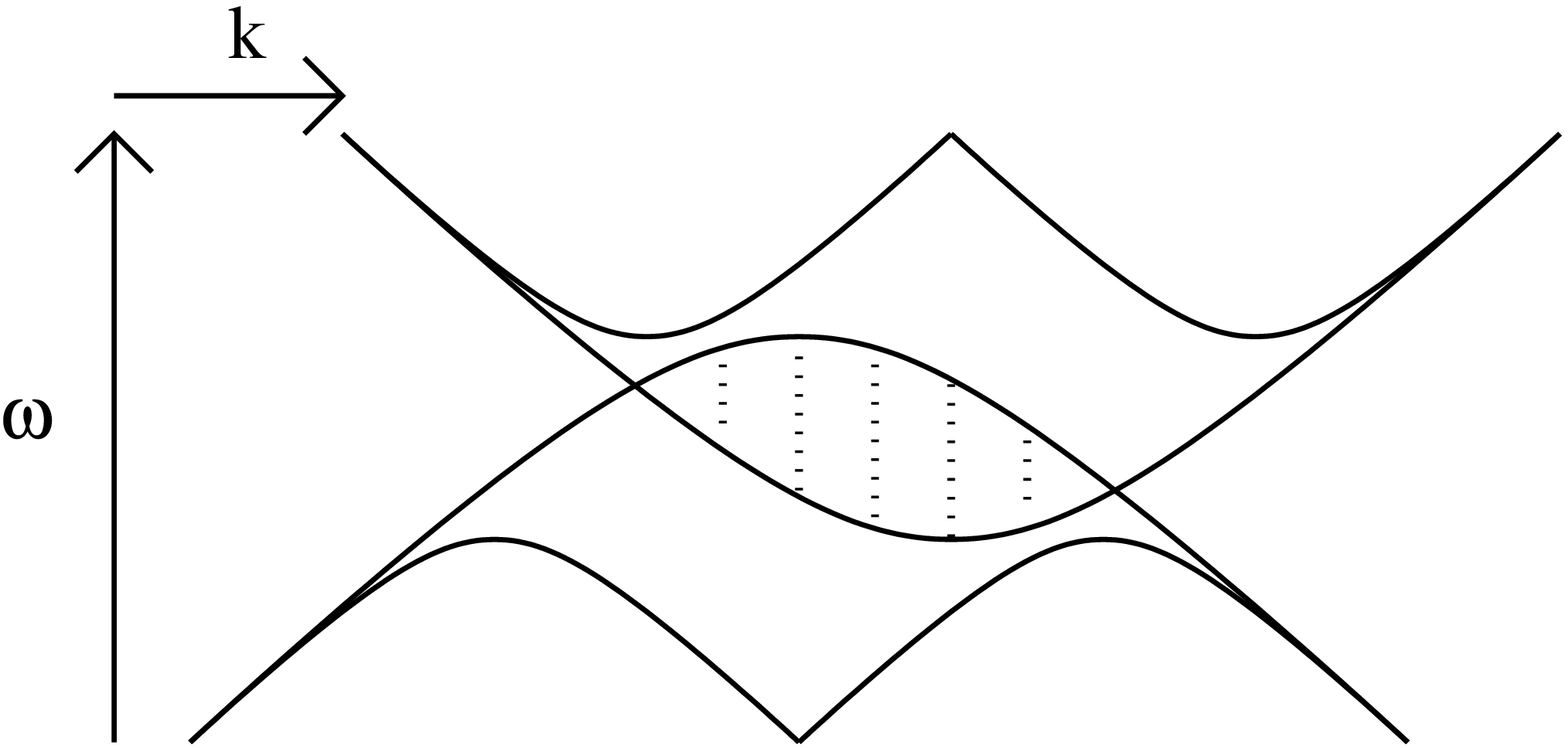}
&
\includegraphics[width=6cm]{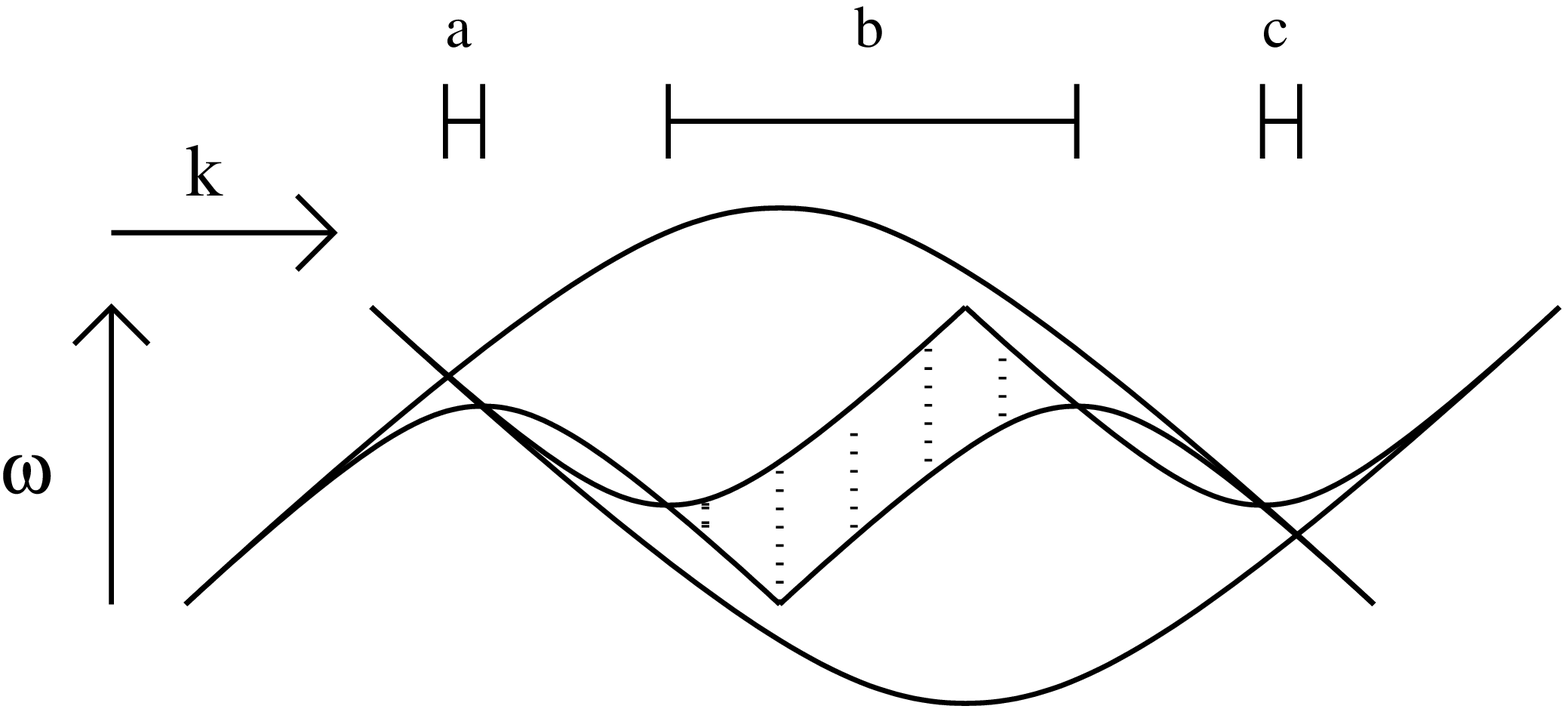}
\end{tabular}
\caption{Integral regions in the $K, \Omega$ plane.  These are formed by the intersection of two 2-spinon
continua, one of them inverted, and shifted with respect to one another by $k$ and $\omega$.  The
left-hand example is a simple case where only one connected domain is obtained.  On the right is a more
complicated example with three different integration domains (whose $K$ borders are pointed out above;  the
domains delimited by $a$ and $c$ have a very small but finite area).}
\label{KW_int_reg}
\end{figure}

\begin{SCfigure}
\includegraphics[width=6cm]{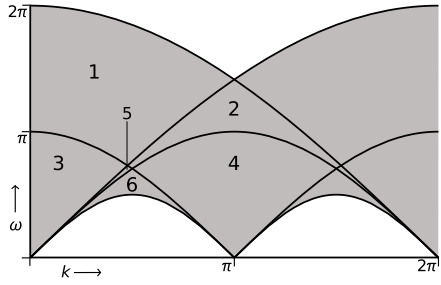}
\label{branch_fig}
\caption{For values of $k$ and $\omega$ within the 4-spinon continuum (shaded), the $K$ and $\Omega$ integration regions
take different forms.  Six sectors are obtained, as labeled here (sector $5$ is barely visible, bordered by sectors 
$2,3,6$).  Examples of $K$, $\Omega$ integration regions
for each of these sectors are given in Figure \ref{calimeros}.}
\end{SCfigure}

For the implementation of the computation of the 4-spinon part of the structure factor, it is desirable to
describe more carefully the actual integration regions for $K$ and $\Omega$, which can be precisely defined for
given $k \in [0, \pi]$ and $\omega \in [ \omega_{4,l} (k), \omega_{4,u}(k)]$.  
Simple reasoning shows that the $K, \Omega$ integration regions are obtained by intersecting two separate 2-spinon continua,
one upright and the other inverted in frequency, and shifted with respect to each
other by $k$ in momentum and $\omega$ in frequency (this is illustrated in Figure \ref{KW_int_reg}).
The important line crossings are respectively of the $\omega_{2,u}$ lines (upper boundaries) of the two continua,
and of their $\omega_{2,l}$ lines.  The first determine which intervals of $K$ should be included,
and the second determine which sub-intervals within these should be excluded.

Regions of $K$ to be included depend on the values of $k$ and $\omega$, and are given by
\begin{eqnarray}
&\omega \leq \pi \sin \frac{k}{2}:  \hspace{2cm} &K \in [0, 2\pi] \nonumber \\
&\pi \sin \frac{k}{2} < \omega \leq 2\pi \sin \frac{k}{4}: &K \in [K_{1a}^-, K_{1a}^+] \cup [K_{1b}^-, K_{1b}^+]
\end{eqnarray}
where 
\begin{eqnarray}
K_{1a}^{\pm} = \frac{k}{2} + \pi \pm 2 \mbox{acos} \frac{\omega}{2\pi \cos\frac{k}{4}}, \hspace{0.5cm}
K_{1b}^{\pm} = \frac{k}{2} \pm 2 \mbox{acos} \frac{\omega}{2\pi \sin\frac{k}{4}}.
\end{eqnarray}
The lower boundaries of the intersecting continua define excluded regions of $K$ as
\begin{eqnarray}
&\frac{\pi}{2} \sin k \leq \omega \leq \pi \sin\frac{k}{2}: \hspace{0.5cm}
&K \notin [K_{2c}^-, K_{2c}^+] \cup [K_{2c}^- + \pi, K_{2c}^+ + \pi], \nonumber \\
&\frac{\pi}{2} \sin k \leq \omega \leq \pi \cos \frac{k}{2}: 
&K \notin [K_{2d}^-, K_{2d}^+] \cup [K_{2d}^- + \pi, K_{2d}^+ + \pi]
\end{eqnarray}
where
\begin{eqnarray}
K_{2c}^{\pm} = \frac{k}{2} + \frac{\pi}{2} \pm \mbox{acos} \frac{\omega}{\pi \cos \frac{k}{2}}, 
\hspace{0.5cm}
K_{2d}^{\pm} = \frac{k}{2} \pm \mbox{acos} \frac{\omega}{\pi \sin \frac{k}{2}}.
\end{eqnarray}
We define the $K$ integration domain ${\cal D}_K$ as the sum of (possibly disconnected)
regions fulfilling the above constraints.

\begin{figure}
\begin{tabular}{ccc}
\includegraphics[width=4cm]{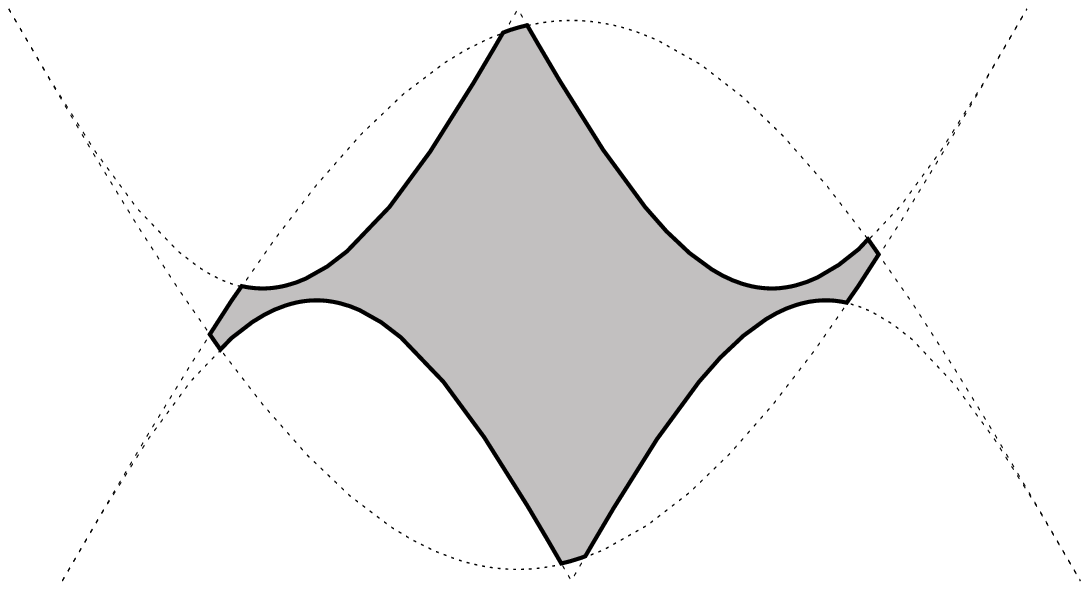}
&
\includegraphics[width=4cm]{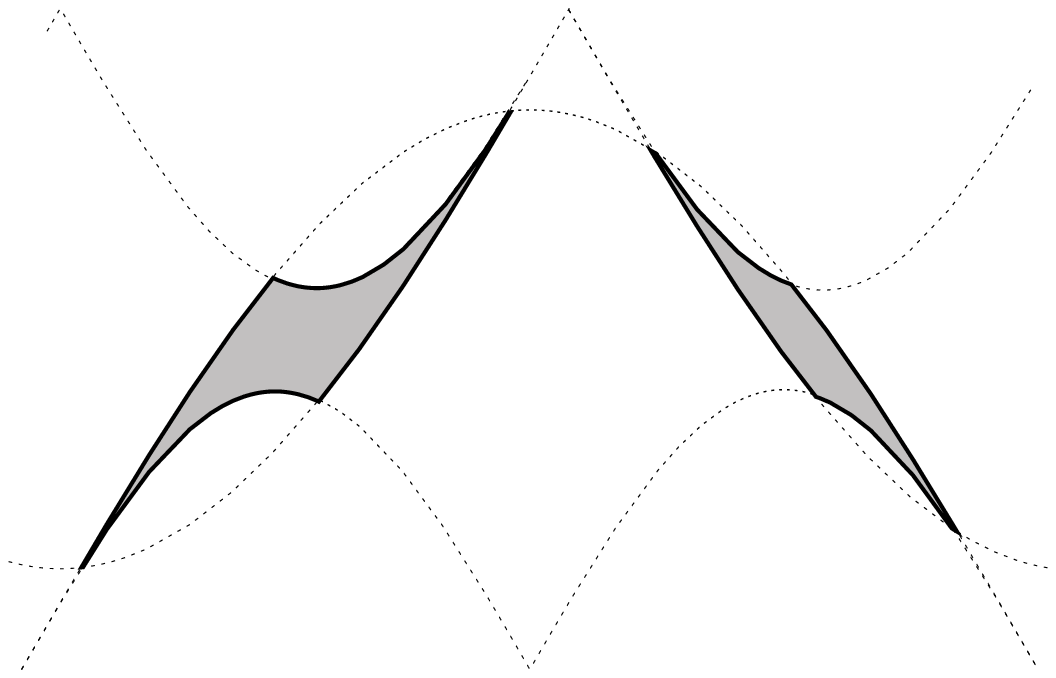}
&
\includegraphics[width=4cm]{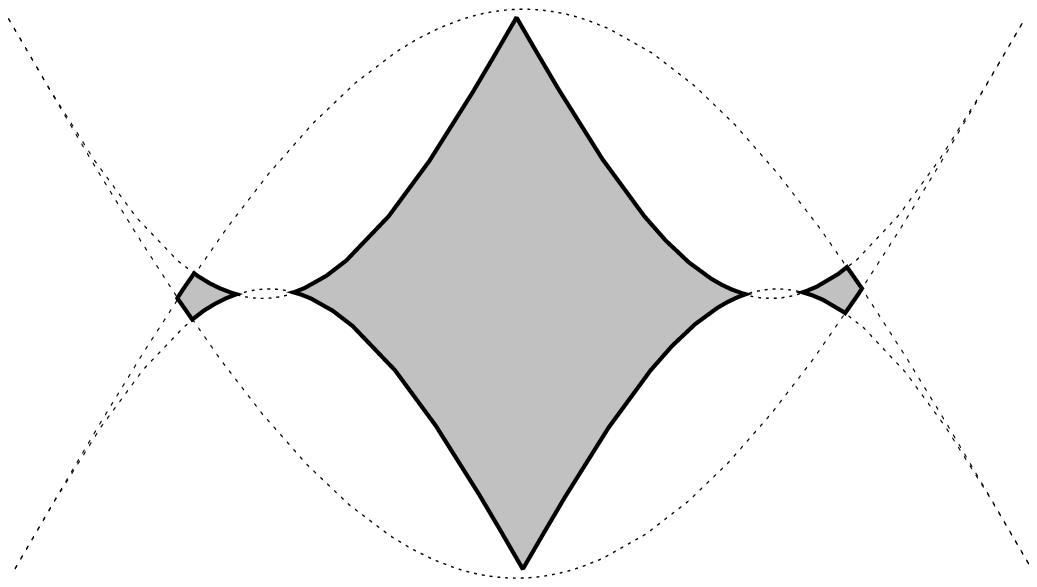} \\
\includegraphics[width=4cm]{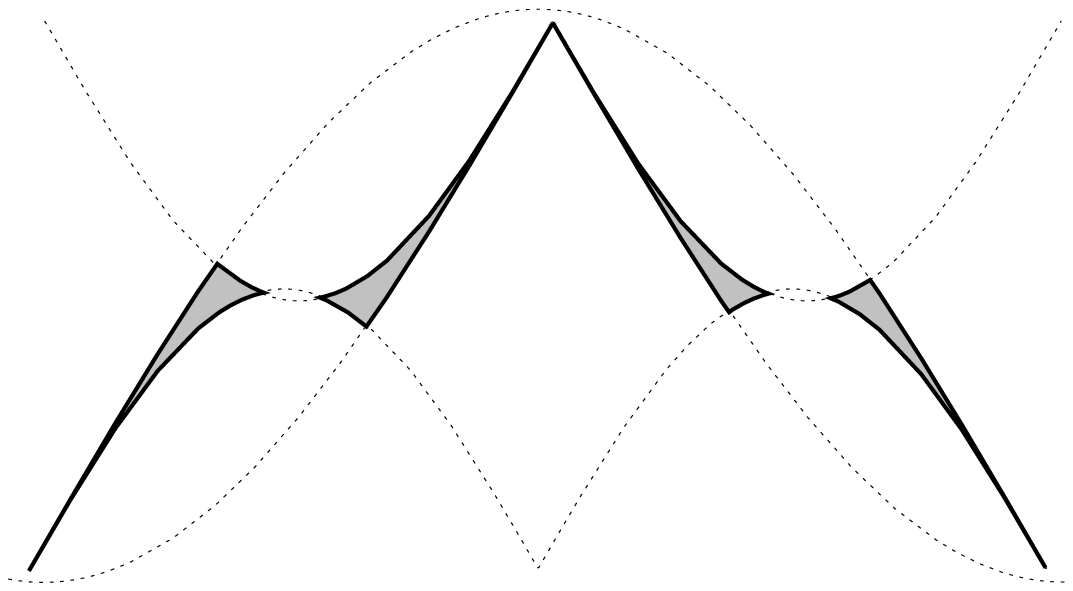} 
&
\includegraphics[width=4cm]{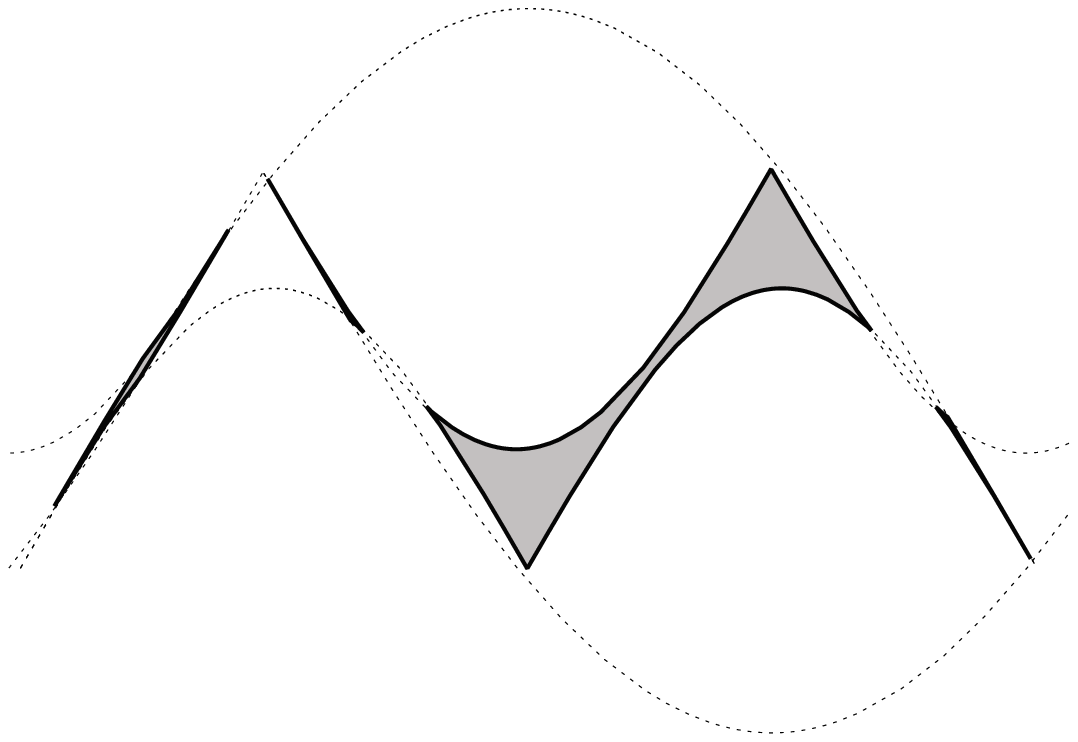}
&
\includegraphics[width=4cm]{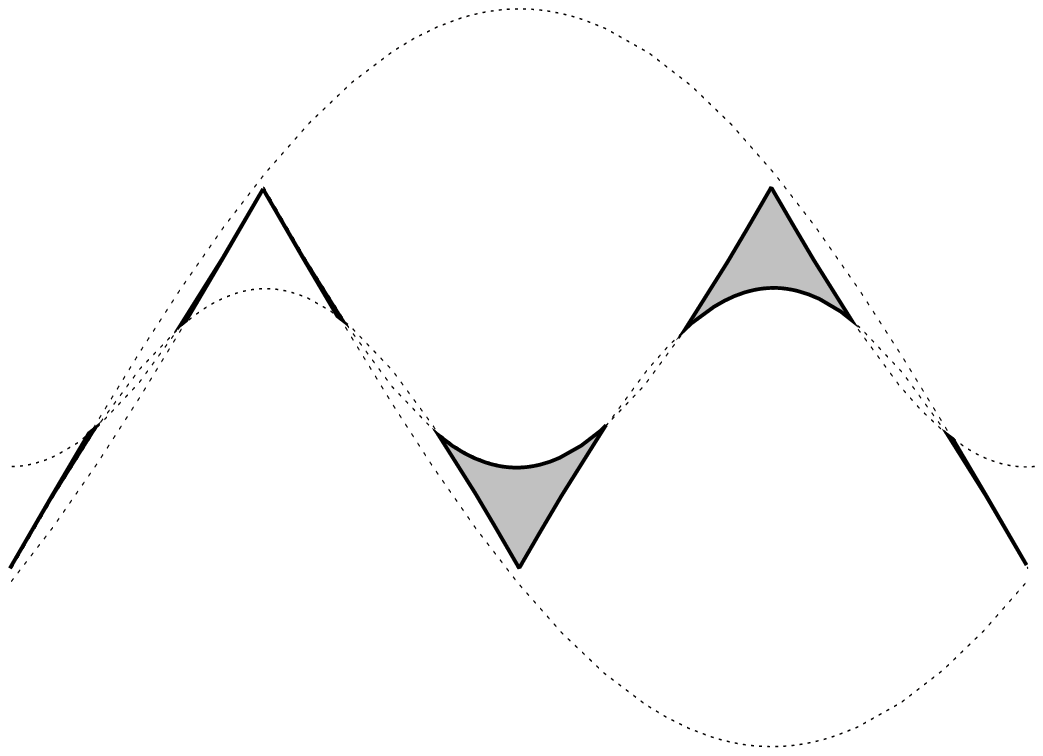}
\end{tabular}
\label{calimeros}
\caption{Integration domains in the $K$, $\Omega$ plane (shaded) for the six sectors given in 
Figure \ref{branch_fig} ($1-3$ top left to right, $4-6$ bottom).  The $K = 0, \Omega = 0$ origin always lies at the left foot
of the upright 2-spinon continuum.}
\end{figure}

For fixed $k, \omega, K$ fulfilling the above constraints, the value of $\Omega$ is restricted to a finite interval:
\begin{equation}
\Omega_{l} (k, \omega, K) \leq \Omega \leq \Omega_u (k, \omega, K),
\end{equation}
with the limits being explicitly given by
\begin{eqnarray}
&&\Omega_l (k, \omega, K) = \mbox{Max} \left(\frac{\pi}{2} | \sin K|, \omega - \pi \sin |\frac{k - K}{2}| \right), \nonumber \\
&&\Omega_u (k, \omega, K) = \mbox{Min} \left(\pi \sin \frac{K}{2}, \omega - \frac{\pi}{2} |\sin (k - K)| \right).
\end{eqnarray}
The leftover two-dimensional integral for the 4-spinon part of the DSF is therefore over a region with
nontrivial geometry, depending on the particular values of $k$ and $\omega$.  Within the 4-spinon continuum,
we can identify six sectors (illustrated in Figure \ref{branch_fig}), each of which leads to a different 
sort of integration domain in the $K$, $\Omega$ plane (examples of which are illustrated in Figure \ref{calimeros}).
By symmetry, we only need to consider $k \in [0, \pi]$.

In terms of the new variables $k, \omega, K, \Omega$, the kinematic restrictions are trivially implemented,
and the four-spinon contribution to the structure factor can be written as a two-dimensional integral
in $K$ and $\Omega$ over the regions defined above:
\begin{eqnarray}
\fl
S_4 (k, \omega) = C_4 \int_{{\cal D}_K} \!\!\!dK \int_{\Omega_l (k, \omega, K)}^{\Omega_u (k, \omega, K)} \!\!\!d\Omega
~\frac{J (k, \omega, K, \Omega)}{\left\{\left[\omega_{2,u}^2 (K) - \Omega^2\right] \left[\omega_{2,u}^2 (k - K) - (\omega - \Omega)^2\right] \right\}^{1/2}}.
\end{eqnarray}
where we have written $J$ implicitly as a function of the new variables.

For the evaluation of this expression, we use a specially adapted Romberg-like integration method.
The integrand typically has integrable divergences at the boundaries of the integration regions,
which are absorbed by various appropriate changes of variables before coding.
While technically feasible as we demonstrate here, it remains a challenge to obtain very 
good precision, and we therefore here first concentrate on results for fixed momentum.

\begin{figure}
\begin{tabular}{cc}
\includegraphics[width=6cm]{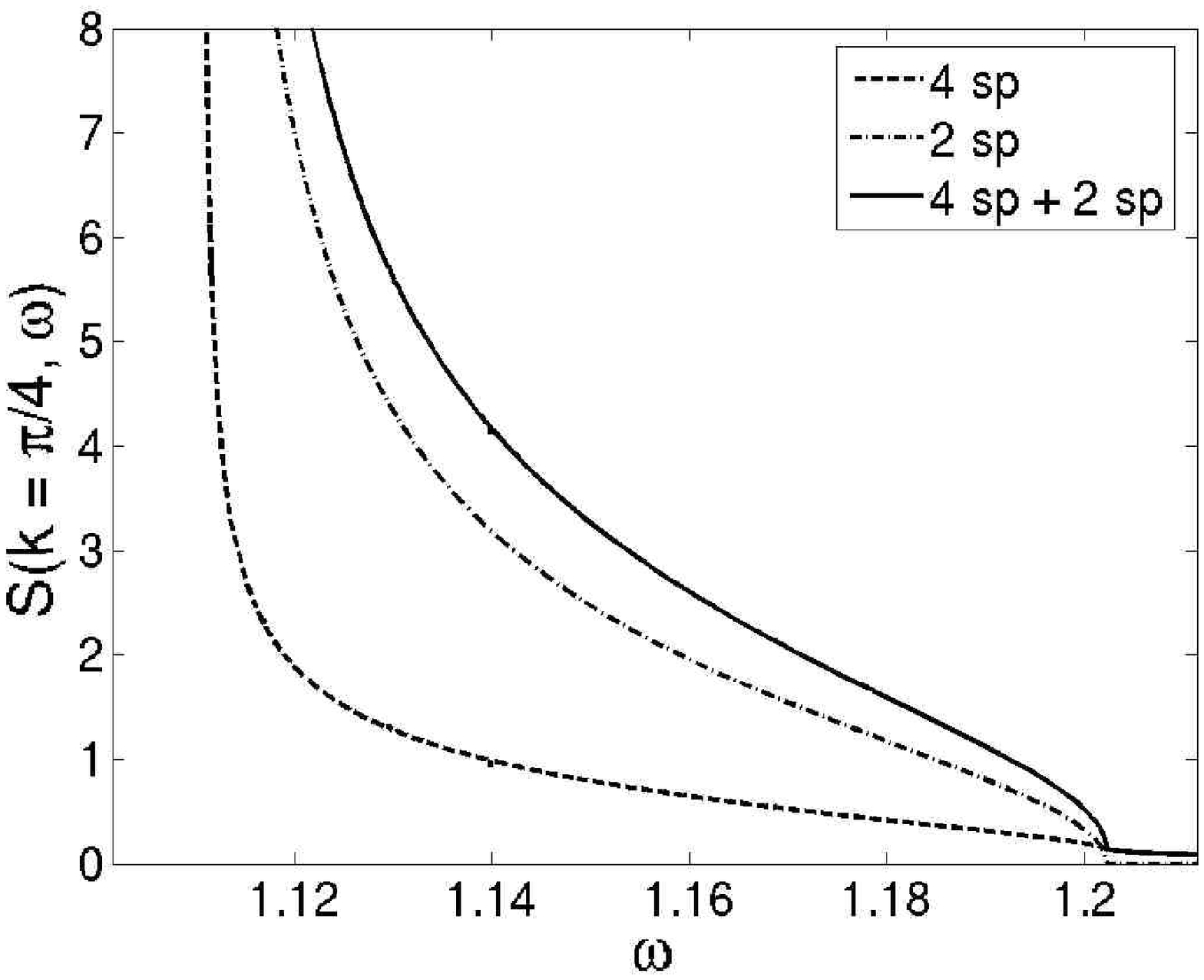}
& 
\includegraphics[width=6cm]{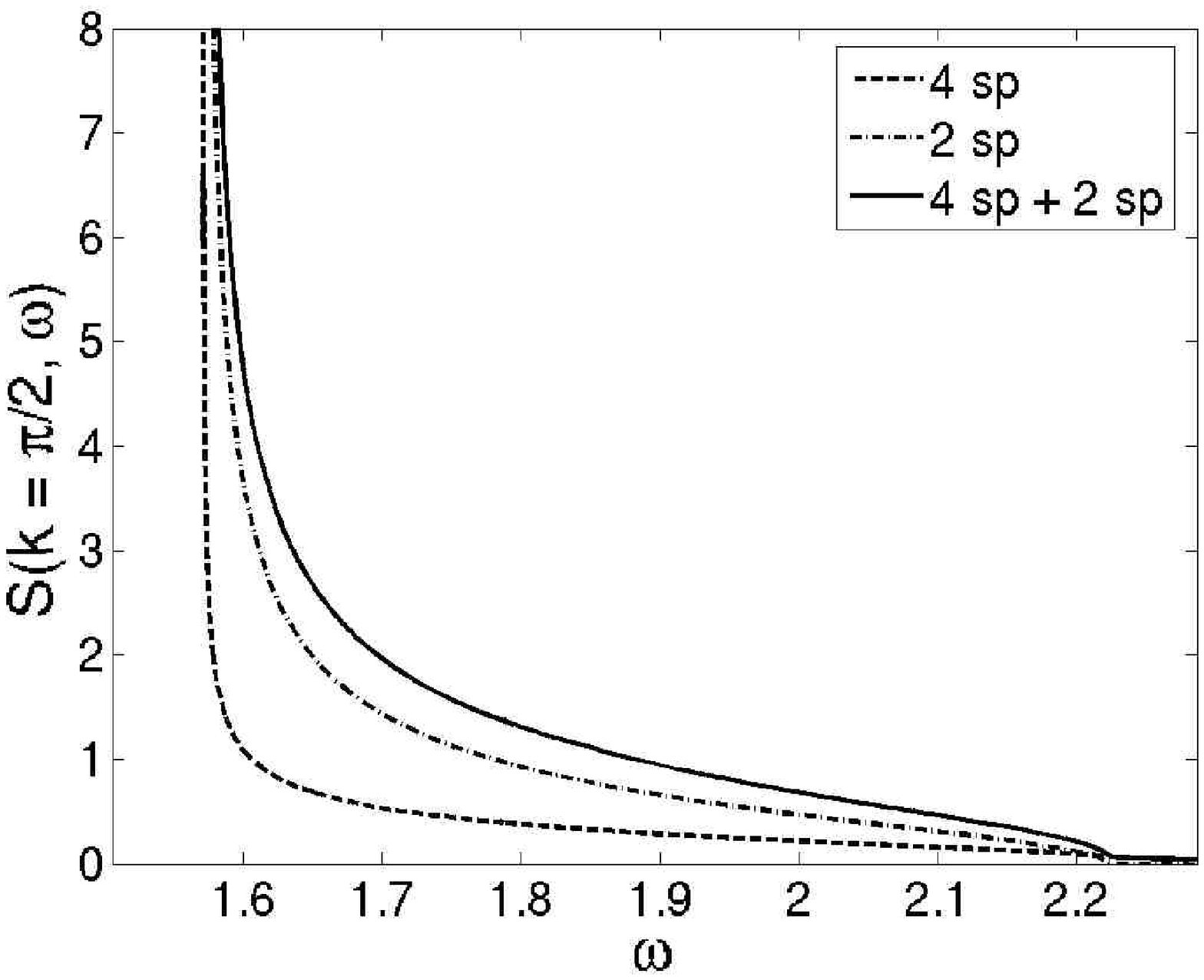} \\
\includegraphics[width=6cm]{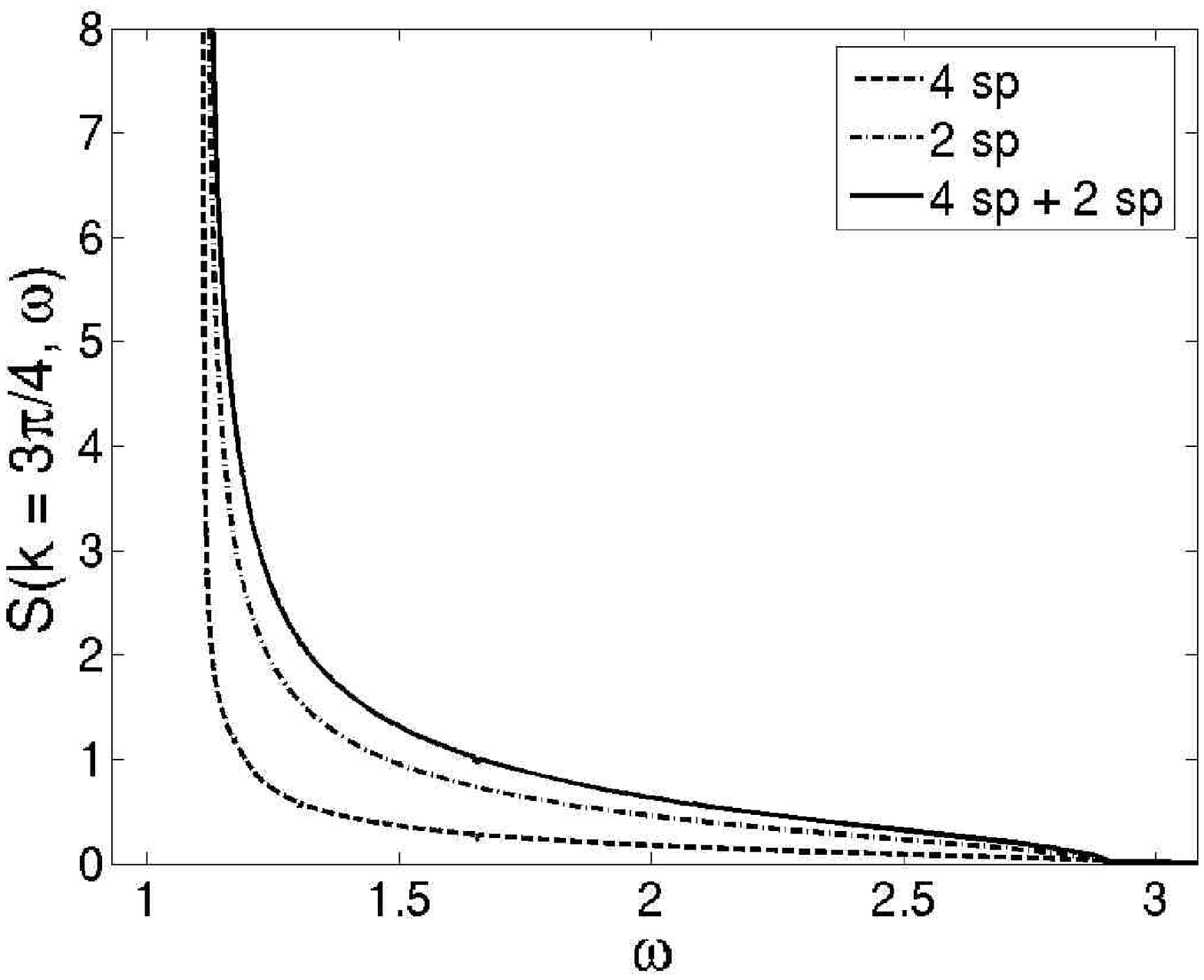}
& 
\includegraphics[width=6cm]{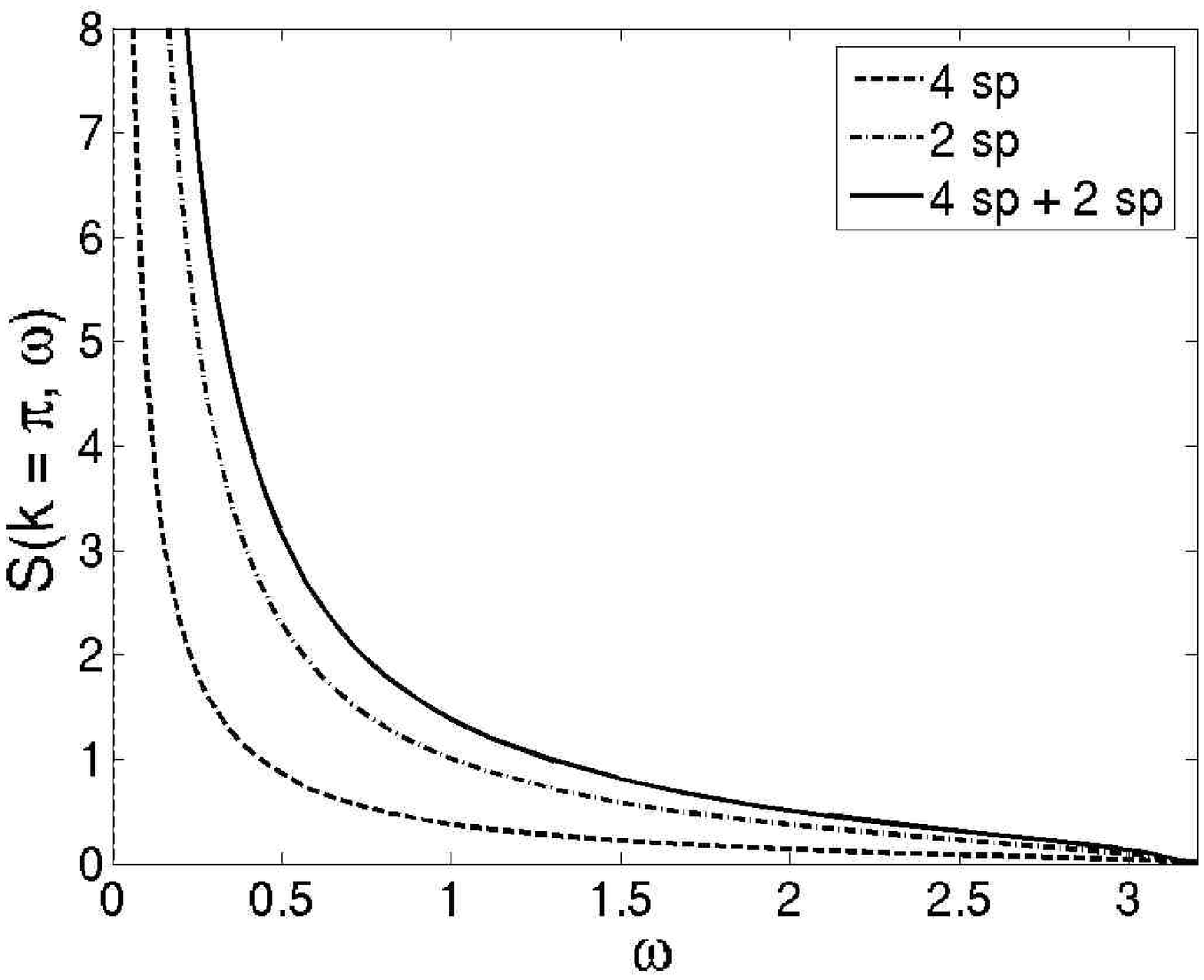} 
\end{tabular}
\label{SF_4p_plots}
\caption{Plots of the 2- and 4-spinon parts of the dynamical structure factor at momenta $k = \pi/4, \pi/2, 3\pi/4$ and $\pi$.
The horizontal axis is frequency, and we focus on the region of the 2-spinon continuum for ease of comparison.}
\end{figure}

The main results are plotted in Figure \ref{SF_4p_plots},
for four representative values of $k$ (other values give similar-looking plots).  
Over the 2-spinon continuum, 
the 4-spinon contribution is of the same order as the 2-spinon one ({\it i.e.} about a third of it).  
Between the upper boundary $\omega_{2,u}$ of the 2-spinon continuum and the upper boundary $\omega_{4,u}$ of
the 4-spinon continuum, the 4-spinon part of the structure factor is finite but very small.  
Figure 5 displays the shape of the DSF in the vicinity of the 2-spinon upper boundary for
two representative values of momentum (the same sort of behaviour is observed for all momentum values
we checked).  A few things are worth pointing out here.  First, at the lower boundary, the 4-spinon
part diverges similarly to the 2-spinon one.  Second, the 4-spinon part is finite and smooth
around the upper boundary $\omega_{2,u}$ of the 2-spinon continuum.  The full DSF therefore 
still has a square-root singularity around this point, yielding a picture consistent with
that put forward in \cite{PustilnikPRL96} (see also the related discussion in \cite{PereiraPRL96}).  
At higher frequencies, the DSF decays extremely rapidly, as illustrated in Figure \ref{SF_4p_higher_omega}.  

\begin{figure}
\begin{tabular}{cc}
\includegraphics[width=6cm]{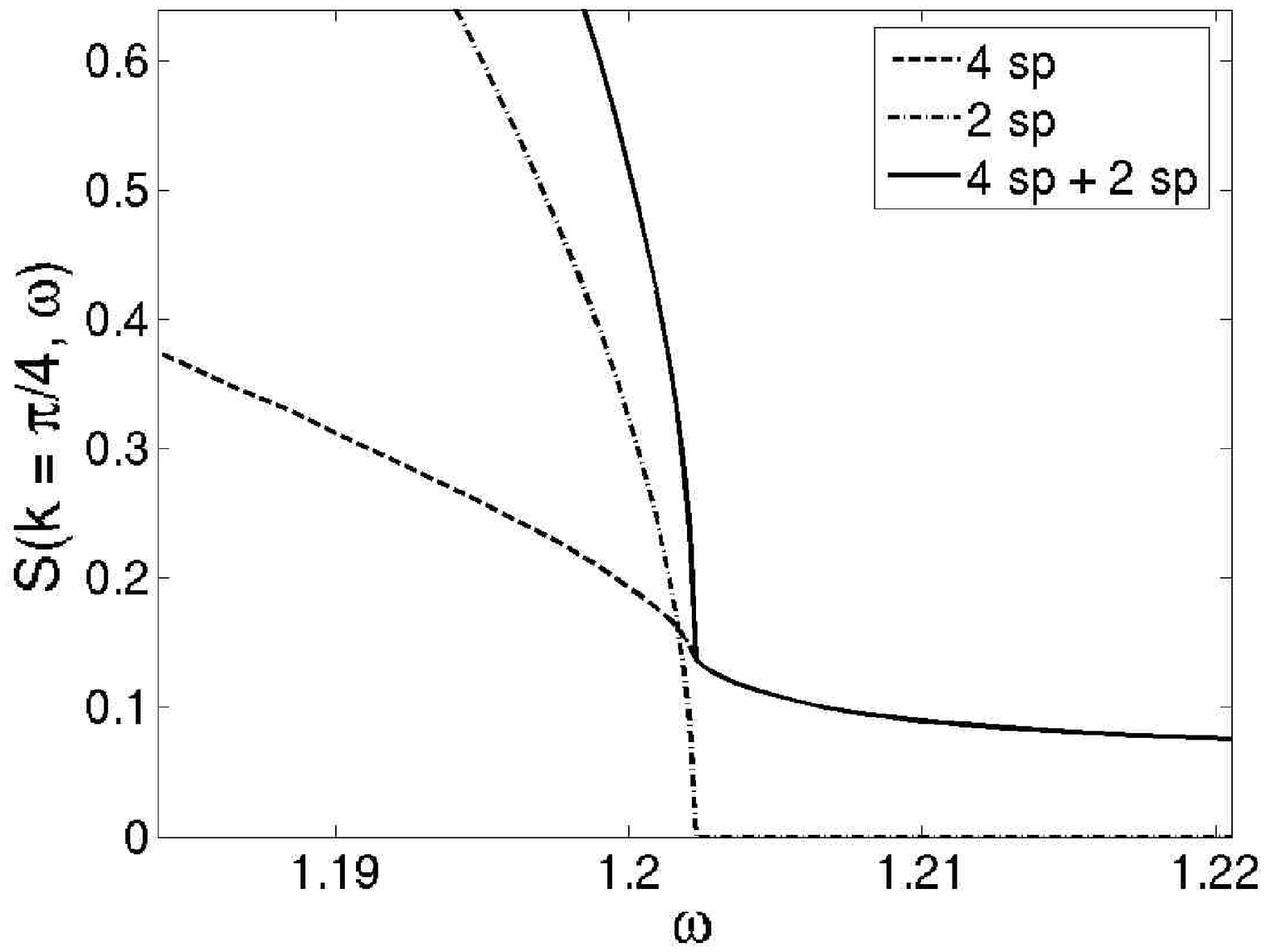}
&
\includegraphics[width=6cm]{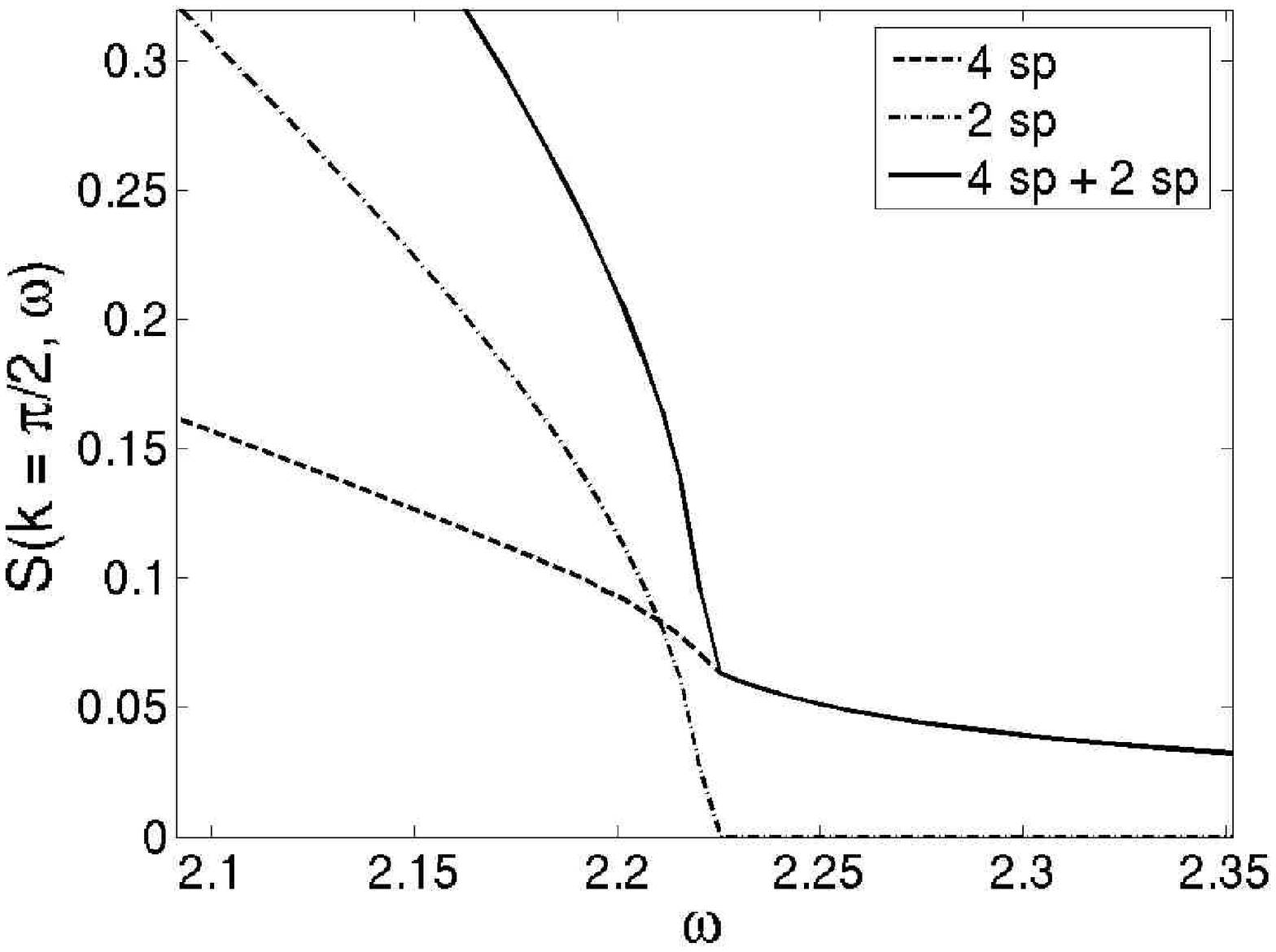}
\end{tabular}
\label{SF_u}
\caption{Zooms on the upper boundary of the 2-spinon continuum, here for $k = \pi/4$ and $\pi/2$ (other values of momentum give
very similar results).  The 2-spinon part vanishes in a square root cusp at $\omega_{2,u}$, whereas the 4-spinon part is 
finite.  There is thus an infinite slope in the full dynamical structure factor at $\omega_{2,u}$.}
\end{figure}

To assess the quality of our results at fixed momentum, we compute the first frequency moment
sum rule in Table 1.  4-spinon intermediate states clearly carry (as expected) the majority
of the missing correlation weight after 2-spinon contributions have been taken into account, 
{\it i.e.} around $27\%$ $\pm 1\%$ of this sum rule for all values of momentum which were studied.  
We have not yet achieved sufficient accuracy for the 4-spinon
part of the total integrated intensity, but we can expect again a contribution of the order of $27\%$
(since the general shape of the 4-spinon part resembles that of the 2-spinon part, and since
the relative 2-spinon part of both sum rules is almost equal).  There thus remains only
about $2\%$ missing, which are naturally ascribed to higher spinon numbers.  Although these are
also in principle accessible using an extension of the present method, the (for 6 spinons, quadruple) integrations needed
probably prohibit accurate evaluation without first achieving significant further analytical advances.
However, this missing part is now rather small, meaning that our results provide an
approximation of the exact zero-temperature DSF of the Heisenberg model in zero field which is the most accurate 
available at the moment.

\begin{SCfigure}
\includegraphics[width=6cm]{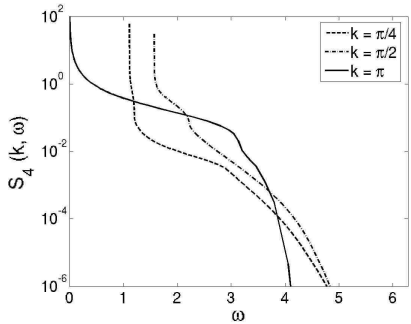}
\label{SF_4p_higher_omega}
\caption{The 4-spinon part (in logarithmic scale) of the DSF as a function of frequency for three representative values of
momentum.  The upper boundary of the 2-spinon continuum is visible as the shoulder in the curve,
with the 4-spinon part of the DSF decreasing rapidly above this threshold.}
\end{SCfigure}

\begin{table}
\begin{tabular}{|c|c|c|c|c|c|c|c|c|}
\hline
$k$ & $\pi/8$ & $\pi/4$ & $3\pi/8$ & $\pi/2$ & $5\pi/8$ & $3\pi/4$ & $7\pi/8$ & $\pi$ \\ \hline \hline 
$K_1 (k)$ (4 sp \%) & 27.2 & 27.3 & 27.3 & 27.1 & 26.5 & 26.2 & 25.9 & 26.6 \\ \hline
\end{tabular}
\caption{First frequency moment sum rule (see equation (\ref{K_1})) percentages coming from the 4-spinon part of the dynamical structure factor.
The 2-spinon part always contributes $71.30\%$.  4-spinon states clearly carry the majority of the leftover correlation
weight, as is naturally expected.  The accuracy of these results is estimated to be around $1\%$ of $K_1(k)$.}
\end{table}

\section*{Conclusion}
In conclusion, we have calculated the 4-spinon contribution to the zero-temperature dynamical structure factor of the 
Heisenberg model in zero magnetic field, starting from its exact integral representation in the thermodynamic limit.
As proven by sum rules, this contribution carries most of the correlation weight left over after 2-spinon intermediate states have been
taken into accout.  The results obtained therefore provide a very close description of the exact correlator.
In future publications, we will provide a thorough analysis of other available sum rules, and extend the results
to the gapped antiferromagnetic case.

\section*{Acknowledgments}
J.-S. C. acknowledges interesting discussions with participants of the 2006 Lyon EUCLID network conference,
in particular with R. Weston, and acknowledges support from the Stichting voor Fundamenteel Onderzoek der Materie 
(FOM) in the Netherlands.


\vspace{0.5cm}

\begin{thebibliography}{99}
\bibitem{BetheZP71} H. Bethe, Z. Phys. 71, 205 (1931).
\bibitem{HeisenbergZP49} W. Heisenberg, Z. Phys. 49, 619 (1928).
\bibitem{KorepinBOOK} V. E. Korepin, N. M. Bogoliubov and A. G. Izergin, ``Quantum Inverse Scattering
Method and Correlation Functions'', Cambridge University Press, 1993, and references therein.
\bibitem{TakahashiBOOK} M. Takahashi, ``Thermodynamics of One-Dimensional Solvable Models'',
Cambridge, 1999, and references therein.
\bibitem{EsslerREVIEW} F. H. L. Essler and R. M. Konik, in {\it Ian Kogan Memorial Volume}, World Scientific, 
Singapore, 2005.
\bibitem{HulthenAMAFA11} L. Hulth\'en, Arkiv Mat. Astron. Fys. A11 26, 1 (1938).
\bibitem{desCloizeauxPR128} J. des Cloiseaux and J. J. Pearson, Phys. Rev. 128, 2131 (1962).
\bibitem{GriffithsPR133} R. B. Griffiths, Phys. Rev. 133, A768 (1964).
\bibitem{YangPR150} C. N. Yang and C. P. Yang, Phys. Rev. 150, 321 (1966); 150, 327 (1966); 151, 258 (1966).
\bibitem{GaudinPRL26} M. Gaudin, Phys. Rev. Lett. 26, 1301 (1971).
\bibitem{TakahashiPTP46_1} M. Takahashi, Prog. Theor. Phys. 46, 401 (1971).
\bibitem{KenzelmannPRB65} M. Kenzelmann, R. Coldea, D. A. Tennant, D. Visser, M. Hofmann, P. Smeibidl and Z. Tylczynski, 
Phys. Rev. {\bf B} 65, 144432 (2002).
\bibitem{StonePRL91} M. B. Stone, D. H. Reich, C. Broholm, K. Lefmann, C. Rischel, C. P. Landee and M. M. Turnbull,
Phys. Rev. Lett. 91, 037205 (2003).
\bibitem{ZaliznyakPRL93} I. A. Zaliznyak, H. Woo, T. G. Perring, C. L. Broholm, C. D. Frost and H. Takagi, Phys. Rev. Lett. 93, 087202 (2004).
\bibitem{LakeNature4} B. Lake, D. A. Tennant, C. D. Frost and S. E. Nagler, Nature Materials 4, 329 (2005). 
\bibitem{FaddeevPLA85} L. D. Faddeev and L. A. Takhtadjan, Phys. Lett. A 85, 375 (1981).
\bibitem{FowlerPRB18} M. Fowler, Phys. Rev. B 18, 421 (1978).
\bibitem{AndersonSCIENCE235} P. W. Anderson, Science 235, 1196 (1987). 
\bibitem{MuellerPRB24} G. M\"uller, H. Thomas, H. Beck and J. C. Bonner, Phys. Rev. B 24, 1429 (1981).
\bibitem{LutherPRB9} A. Luther and I. Peschel, Phys. Rev. B 9, 2911 (1974); {\it ibid}. 12, 3908 (1975).
\bibitem{KadanoffAP121} L. P. Kadanoff and A. C. Brown, Ann. Phys. (N.Y.) 121, 318 (1979). 
\bibitem{AffleckPRL56} I. Affleck, Phys. Rev. Lett. 56, 746 (1986).
\bibitem{BloetePRL56} H. W. J. Bl\"ote, J. L. Cardy and M. P. Nightingale, Phys. Rev. Lett. 56, 742 (1986).
\bibitem{BelavinNPB241} A. A. Belavin, A. M. Polyakov and A. B. Zamolodchikov, Nucl. Phys. B 241, 333 (1984).
\bibitem{CFTBOOK} P. Di Francesco, P. Mathieu and D. S\'en\'echal, {\it Conformal Field Theory}, Springer, 1997, and references therein.
\bibitem{AffleckLESHOUCHES} I. Affleck, in {\it Les Houches, session XLIX, Fields, strings and critical phenomena},
Elsevier, New York, 1989.
\bibitem{BosonizationBOOK} A. O. Gogolin, A. A. Nersesyan and A. M. Tsvelik, {\it Bosonization and Strongly
Correlated Systems}, Cambridge University Press, Cambridge, 1998.
\bibitem{GiamarchiBOOK} T. Giamarchi, {\it Quantum Physics in One Dimension}, Oxford University Press, Oxford, 2004.
\bibitem{AffleckJPA31} I. Affleck, J. Phys. A 31, 4573 (1998).
\bibitem{LukyanovPRB59} S. Lukyanov, Phys. Rev. B 59, 11163 (1999).
\bibitem{LukyanovNPB654} S. Lukyanov and V. Terras, Nucl. Phys. B 654, 323 (2003).
\bibitem{CauxPRL95} J.-S. Caux and J. M. Maillet, Phys. Rev. Lett. {\bf 95}, 077201 (2005).
\bibitem{CauxJSTATP09003} J.-S. Caux, R. Hagemans and J. M. Maillet, J. Stat. Mech. (2005) P09003.
\bibitem{ABACUS} Algebraic Bethe Ansatz Computation of Universal Structure factors.  
See {\sf http://staff.science.uva.nl/$\sim$jcaux/ABACUS.html}. 
\bibitem{KitanineNPB554} N. Kitanine, J. M. Maillet and V. Terras, Nucl. Phys. B 554, 647 (1999).
\bibitem{KitanineNPB567} N. Kitanine, J. M. Maillet and V. Terras, Nucl. Phys. B 567, 554 (2000).
\bibitem{DaviesCMP151} O. Davies, O. Foda, M. Jimbo, T. Miwa and A. Nakayashiki, Commun. Math. Phys. 151, 89 (1993).
\bibitem{FrenkelPNAS85} I. B. Frenkel and N. H. Jing, Proc. Natl. Acad. Sci. 85, 9373 (1988).
\bibitem{AbadaMPLA8} A. Abada, A. H. Bougourzi and M. A. El Gradechi, Mod. Phys. Lett. A 8, 715 (1993).
\bibitem{BougourziNPB404} A. H. Bougourzi, Nucl. Phys. B 404, 457 (1993).
\bibitem{JimboBOOK} M. Jimbo and T. Miwa, ``Algebraic Analysis of Solvable Lattice Models'', AMS, Providence (1995).
\bibitem{BougourziPRB54} A. H. Bougourzi, M. Couture and M. Kacir, Phys. Rev. B 54, R12669 (1996).
\bibitem{BougourziPRB57} A. H. Bougourzi, M. Karbach and G. M\"uller, Phys. Rev. B 57, 11429 (1998).
\bibitem{KarbachPRB55} M. Karbach, G. M\"uller, A. H. Bougourzi, A. Fledderjohann and K.-H. M\"utter, Phys. Rev. B 55, 12510 (1997).
\bibitem{WestonCRM} R. A. Weston and A. H. Bougourzi, Preprint CRM-2198, 1994 (unpublished).
\bibitem{BougourziMPLB10} A. H. Bougourzi, Mod. Phys. Lett. B 10, 1237 (1996).
\bibitem{AbadaNPB497} A. Abada, A. H. Bougourzi and B. Si-Lakhal, Nucl. Phys. B 497, 733 (1997).
\bibitem{Abada9802271} A. Abada, A.H. Bougourzi, S. Seba, B. Si-Lakhal, cond-mat/9802271.
\bibitem{AbadaJPA37} A. Abada and B. Si-Lakhal, J. Phys. A: Math. Gen. 37, 497 (2004).
\bibitem{SiLakhalPB369} B. Si-Lakhal and A. Abada, Physica B 369, 196 (2005).
\bibitem{PustilnikPRL96} M. Pustilnik, M. Khodas, A. Kamenev and L. I. Glazman, Phys. Rev. Lett. 96, 196405 (2006).
\bibitem{PereiraPRL96} R. G. Pereira, J. Sirker, J.-S. Caux, R. Hagemans, J. M. Maillet, S. R. White and I. Affleck,
Phys. Rev. Lett. 96, 257202 (2006).
\end{thebibliography}
\end{document}